\documentclass{elsart}

\input{psfig.tex}

\newcommand{\eed}{($e$,$\,e'd$)}

\newcommand{\pn} {$pn$}

\newcommand{\hedrie}{$\,^{\rm 3}$He}

\newcommand{\heep}  {$^1$H($e$,$\,e'p$)}
\newcommand{\deed}  {$^2$H($e$,$\,e'd$)}
\newcommand{\hedrieeed}    {$\,^{\rm 3}$He$(e,\,e'd)$}
\newcommand{\hedrieeedp}   {$\,^{\rm 3}$He$(e,\,e'd)p$}

\newcommand{\heviereedd}   {$\,^{\rm 4}$He$(e,\,e'd)d$}
\newcommand{\heviereedpn}  {$\,^{\rm 4}$He$(e,\,e'd)pn$}

\newcommand{\qw}    {($q,\,\omega$)}
\newcommand{\mevc}  {MeV/$c$}

\begin{document}

\begin{frontmatter}

\title{The $^3$He(e,\,e$'$d)p Reaction in \qw-constant Kinematics}

\author[a,b] {C.M.~Spaltro\thanksref{chiara}},
\author[b,c] {Th.S.~Bauer},
\author[a,b] {H.P.~Blok},
\author[b] {T.~Botto},
\author[d] {E.~Cisbani},
\author[e] {R.~De~Leo},
\author[a,b] {G.E.~Dodge},
\author[g] {R.~Ent},
\author[d] {S.~Frullani},
\author[b] {F.~Garibaldi},
\author[h] {W.~Gl\"{o}ckle},
\author[h,i] {J.~Golak},
\author[f] {M.N.~Harakeh},
\author[d] {M.~Iodice}, 
\author[b] {E.~Jans},
\author[j] {H.~Kamada},
\author[b] {W.J.~Kasdorp}, 
\author[a,b] {C.~Kormanyos}, 
\author[b] {L.~Lapik\'{a}s}, 
\author[b,c] {A.~Misiejuk}, 
\author[a,b] {S.I.~Nagorny}, 
\author[b] {G.J.~Nooren}, 
\author[a,b] {C.J.G.~Onderwater}, 
\author[e] {R.~Perrino}, 
\author[a,b] {M.~van~Sambeek}, 
\author[i] {R.~Skibi\'nski},
\author[a,b] {R.~Starink},
\author[b] {G.~van~der~Steenhoven}, 
\author[c] {J.~Tjon},
\author[b] {M.A.~van~Uden}, 
\author[d] {G.M.~Urciuoli}, 
\author[b] {H.~de~Vries},
\author[i] {H.~Wita\l{}a},
\author[b,c] {and M.~Yeomans}
\address [a] {Vrije~Universiteit~Amsterdam, de~Boelelaan~1081, 1081~HV~Amsterdam,
The~Netherlands}
\address [b] {Nationaal Instituut voor Kernfysica en Hoge-Energiefysica
(NIKHEF),
P.O.~Box~41882, 1009~DB~Amsterdam, The~Netherlands}
\address [c] {Universiteit~Utrecht, P.O.~Box~80.000, 3508~TA~~Utrecht, The~Netherlands}
\address [d] {INFN~Sezione~Sanit\'{a}, V.le~Regina~Elena~299, 00161~Roma, Italy}
\address [e] {INFN, Sezione di Lecce, via Arnesano, 73100~Lecce, Italy}
\address [f] {KVI, Rijksuniversiteit~Groningen, Zernikelaan~25, 9747~AA~Groningen,
The~Netherlands}
\address [g] {Thomas Jefferson National Accelerator Facility, Newport~News, VA.~23606,
USA}
\address [h] {Institut f\"{u}r Theoretische Physik II, Ruhr-Universit\"{a}t, 
D-44780 Bochum, Germany}
\address [i] {Institute of Physics, Jagellonian University, PL-30059 Cracow, Poland}
\address [j] {Kyushu Institute of Technology, 1-1 Sensuicho, Tobata,
Kitakyushu 804-8550, Japan}
%
\thanks[chiara]{present address: CMG, Graadt van Roggenweg 350, 3531 AH, Utrecht, The
Netherlands}

\begin{abstract}
The cross section for the  \hedrieeedp\ reaction has been measured
as a function of the missing momentum $p_m$
in \qw -constant kinematics at beam energies of 370 and 576~MeV 
for values of the three-momentum transfer $q$ of 412, 504 and 604~\mevc.
The L(+TT), T and LT structure functions  
have been separated for $q$~=~412 and 504~\mevc.
The data are compared to three-body Faddeev calculations,
including meson-exchange currents (MEC), and to calculations based on a
covariant diagrammatic expansion.
The influence of final-state interactions and meson-exchange currents is
discussed.
The $p_m$-dependence of the data is reasonably well described
by all calculations. However, the most advanced Faddeev calculations,
which employ the AV18 nucleon-nucleon interaction and
include MEC, overestimate the measured cross sections,
especially the longitudinal part, and at the larger values of $q$.
The diagrammatic approach gives a fair description of the cross section,
but under(over)estimates the longitudinal (transverse) structure function.

\end{abstract}

\begin{keyword}

\PACS 21.45+v \sep 25.10.+s \sep 25.30.Dh \sep 25.30.Fj


\end{keyword}

\end{frontmatter}


\section{Introduction}
\label{sect:intro}

Many nuclear properties can be described successfully within a mean-field
approach. However, phenomena like the depletion of spectroscopic strength
and the occurrence of bumps at missing energies characteristic of two-nucleon
emission in $(e,e'p)$ reactions, indicate that correlations between nucleons,
i.e., the motion inside a nucleus of two nucleons as a pair with a certain
relative motion, also play an essential role.
The \eed\ reaction has proven to be a sensitive tool for the investigation of
pro\-ton-neu\-tron (\pn ) correlations in nuclei
\cite{kei85,ent86,ent89,ent91,iod92,ent94,tri96}.
Assuming direct knock-out of a \pn\ pair, the cross section for
the \eed\ reaction can approximately be written  \cite{ent94}
as $d^6\sigma/dE_{e'}d\Omega_{e'}dE_{d}d\Omega_{d}
= K \sigma_{e,pn}(q) S_{pn}(E_m,p_m,p_{d})$.
Here, the cross section $\sigma_{e,pn}$ for scattering of an electron
from a \pn\ pair leading to a deuteron in the final state,
which depends on the momentum transfer $q$,
reflects the relative proton-neutron motion,
i.e., the relative \pn\ wave function.
The distorted spectral function $S_{pn}$,
which depends on the missing energy $E_m$, the missing momentum $p_m$
and the momentum of the final deuteron $p_{d}$,
contains the information about the centre-of-mass (c.o.m.) motion of the
\pn\ pair within the nucleus, modified by final-state interactions (FSI).
Since the deuteron has isospin zero and the initial \pn\ pair can
be in a $T=0$ or $T=1$ state, in general both $\Delta T=0$ and
$\Delta T=1$ transitions are possible. The former resembles quasi-elastic
knockout of a deuteron, whereas the latter is similar to the (inverse of)
deuteron disintegration. Both types of transitions have been studied
previously (\cite{ent86,ent91,iod92,ent94}, \cite{ent89,ent94,tri96}).

In this paper we present data for the \hedrieeed\ reaction, taken in
so-called \qw-constant kinematics, in which the energy and momentum transfer
to the nucleus are held constant, while the angle of the outgoing deuteron
is varied. The data is compared to the results of three-body calculations
\cite{mei90,ish94,gol95}
and to those of a covariant and gauge-invariant diagrammatic approach
\cite{nag89,nag91,nag94a,nag94b}.
For the three-nucleon system Faddeev calculations are nowadays available
both for the ground state and for the continuum.
The first available \hedrieeed\ data \cite{kei85} has been compared to
the results of such calculations \cite{mei90,gol95}.
The emphasis in these calculations was on the correct treatment of FSI,
which turned out to be crucial, especially for kinematics
dominated by deuteron knock-out processes. The same data was also employed
in Refs.~\cite{nag89,nag91,nag94a}, with the emphasis on the consistency
between the dynamics and the one- and many-body currents.
 The agreement with the data, especially for parallel kinematics,
was in both cases fairly good.
Other data \cite{tri96} was of limited kinematic coverage and
no comparison to Faddeev calculations was made.
More recently, we published data for the \hedrieeed\ reaction taken in
parallel kinematics \cite{spa98}. Theoretical calculations gave a fair
description of the data, but there were some discrepancies, especially for
the $q$ dependence, and for the transverse structure function $W_T$.

Data in \qw-constant kinematics  extend the range in which the theoretical
predictions can be tested, and feature some new aspects.
First of all one can investigate the longitudinal-transverse (LT)
 interference structure function,
which is more sensitive to details of the interaction currents.
Furthermore, one can investigate the dependence of the longitudinal (L)
and transverse (T) structure functions on the kinematics.
Finally, in selected parts
of this kinematics one gets non-negligible contributions from
processes in which the photon hits the non-detected proton.
Interference with the process where the photon interacts with
(the nucleons of) the deuteron gives an extra sensitivity
to a correct description of the reaction.

Since the publication of the data obtained in parallel kinematics there
have been several theoretical developments (see section~\ref{sect:theory}).
The Faddeev calculations by Golak \etal\ \cite{gol95}
were extended by using
the AV18 NN potential in addition to the Bonn-B potential,
and by including meson-exchange currents (MEC).  Furthermore,
the number of partial waves was increased to ensure full convergence.
New calculations in the diagrammatic approach by Nagorny
\cite{nag89,nag91,nag94a,nag94b} employed
also the AV18 potential instead of the Reid Soft Core potential, and now
include all angular momentum states in the \hedrie\ ground state.

Since the data to be discussed comprise both the $q$ and $p_m$
dependence of the reaction, different aspects of the \pn\ motion
in the nucleus $^3$He can be studied. Furthermore,
the cross sections were measured at two beam energies, so that a
separation of the longitudinal ($W_L$) plus transverse-transverse ($W_{TT}$),
transverse ($W_T$), and longitudinal-transverse interference ($W_{LT}$)
structure functions could be performed.
\newline
For the definition of the structure functions
we follow Raskin and Donnelly \cite{ras89}, who write
the differential cross section for the unpolarized \eed\ reaction as:
\begin{equation}
\frac{d^5 \sigma}{dE_{e'} d\Omega_{e'} d\Omega_d} =
C\  \bigl( v_L W_L  +  v_T W_T + v_{LT} W_{LT} \cos\phi
+v_{TT} W_{TT} \cos2\phi
\bigr),
\label{eq:xs_in_vW}
\end{equation}
where $\phi$ is the angle between the electron scattering plane
and the plane defined by the momentum transfer $\vec{q}$ and the momentum
of the outgoing deuteron $\vec{p_d}$.
The factor $C$ contains the Mott cross section and kinematical factors.
Its precise form, as well as that of the kinematical factors $v_i$,
are given in Ref.~\cite{ras89}.
The structure functions $W_{LT}$ and $W_{TT}$ are zero in parallel kinematics.

The separate structure functions, which result from different combinations
of the components of the nuclear current, have a different sensitivity to the
various aspects of the reaction. For instance, the coupling of the virtual
photon to a $T=1$ \pn\ pair, transforming it into a deuteron, involves
a spin-flip and thus is purely transverse, whereas the coupling to an
initial $T=0$ pair, which resembles elastic $e-d$ scattering, is
dominantly longitudinal at our values of $q$
(see also \cite{ent89,tri96}).
Furthermore, one expects that MEC will mainly contribute to $W_T$.
On the other hand, according to the three-body calculations the effects of FSI
on $W_L$  are quite different from the ones on $W_T$.


\section{Experiment}
\label{sect:exp}

The \hedrieeedp\ reaction has been measured ``in-plane'',
using \qw -constant kinematics, for missing momenta up to 210~\mevc,
and for values of the transferred momentum $q$ of 412, 504 and 604~\mevc.
The (central) kinematics is given in Table~\ref{tab:kin}.
\begin{table}[hbt]
        \begin{center}
        \caption{
        Overview of the kinematics. $T_{\rm cm}$ denotes the centre-of-mass
        energy of the outgoing deuteron.
        }
        \vspace{0.5cm}
        \begin{tabular}{|*{3}{@{\hspace{5mm}}c@{\hspace{5mm}}}|c|c|}
        \hline
        $q$ & $\omega$ & $T_{\rm cm}$ & $\theta_{e'}$ (576 MeV) &
        $\theta_{e'}$ (370 MeV)\\
        (\mevc) & (MeV) & (MeV)  & (deg.) & (deg.)\\
        \hline
        412 & 50.0  & 14.7 & 43.6 & 72.9 \\
        504 & 70.0  & 21.0 & 55.1 & 97.0 \\
        604 & 100.0 & 31.1 & 69.3 & -    \\
        \hline
        \end{tabular}
        \label{tab:kin}
        \end{center}
\end{table}
In later figures a negative sign of $p_m$ denotes kinematics in which
$\theta_d < \theta_q$ (corresponding to $\phi=0$ in Eq.~\ref{eq:xs_in_vW}),
and a positive $p_m$ kinematics with $\theta_d > \theta_q$
(corresponding to $\phi=\pi$ in Eq.~\ref{eq:xs_in_vW}).
\newline
Since the experiment has been described in detail in
Refs.~\cite{thesis,spa99}, only the main points are mentioned here.

The experiment was  performed with the extracted electron beam from
the pulse-stretcher ring AmPS \cite{wit93} at NIKHEF.
The beam energies were 370 and 576~MeV.
The extracted electron current was about 6~$\mu$A and had a duty factor
of about 70\%.
The scattered electrons were detected with the QDD spectrometer and
the knocked-out deuterons with the QDQ spectrometer \cite{vri84}.

A cryogenic gas target \cite{una95} operating at 20~K and 1.5~Mpa was used,
which was filled with a mixture of $^3$He and $^4$He gases.
In this way data was collected simultaneously for the three reactions:
\hedrieeedp , \heviereedd\ and \heviereedpn .
Results on the latter reaction channel have been published
separately \cite{spa99}.

The resolution in missing energy varied between 0.3 and 2.0~MeV,
depending on the kinematics,
which was sufficient to separate the different reaction channels.
The absolute $^3$He and $^4$He target thicknesses were determined by
comparison of the measured elastic scattering cross sections to
calculated ones \cite{amr94,vri87}.
During the \eed\ measurements the total target thickness was
monitored through the singles rate of either one of the spectrometers.
Checks with elastic scattering before and after each
set of \eed\ measurements were consistent to within 2\%.

Both the \deed\ and the \heep\ reaction were used in order to check
the reconstruction of the electron and hadron momentum vectors
in the QDD and QDQ spectrometers and to determine
the coincidence detection efficiency.

The data analysis included the following steps.
First, the deuterons were separated from protons and tritons
by using the pulse height from the scintillators in the QDQ spectrometer.
Next, the particle vectors at the target were reconstructed.
Since an extended target was used, this reconstruction and the
acceptances of both spectrometers depend on the position of the interaction
point along the beam.
By using energy and momentum conservation the values of the missing energy
\begin{equation}
E_m=\omega-T_p-T_d,
\end{equation}
where $T_x$ is the kinetic energy of particle $x$, and the missing momentum
\begin{equation}
p_m=\vec{q}-\vec{p_d}
\end{equation}
were calculated from the particle vectors.
Next, the accidental coincidences were subtracted.
As a result of the high duty-factor of the extracted beam,
the real-to-accidental ratio was high,
ranging from 15 to 1640 depending on the kinematics.
Next, the data was normalised to the target thickness,
the integrated charge and the experimental detection volume,
and corrected for detection inefficiencies.
The detection volume was obtained from a Monte-Carlo simulation,
which uses the measured optical properties of the
spectrometers~\cite{blo87,off87,vri90}, including their
vertex-position dependent angular acceptances.
Finally, the data was radiatively unfolded.
The systematic uncertainty amounts to about 3\% for the cross sections
and 4-5\% for the structure functions~\cite{thesis,spa99}.

The cross section measured in one kinematical setting
covers an appreciable range in $p_m$,
as a consequence of the angular and momentum  acceptances
of the spectrometers.
However, over this range the values of $q$ and
the kinematical factors in the cross-section expression of
Eq.~(\ref{eq:xs_in_vW})  vary slightly around the central values.
A Monte-Carlo simulation was performed to determine
the average values of the kinematical factors and of $q$
for the different $p_m$ bins within one measurement.
Then, the cross sections for the different $p_m$ bins were
recalculated to a common $q$ value by using the $q$ dependence of
the cross section as measured in this experiment.
The correction to the cross section amounted to  typically
a few percent.


\section{Theoretical calculations}
\label{sect:theory}

The experimental data is compared to the calculations of
Van Meijgaard and Tjon \cite{mei90}, Golak \etal\ \cite{ish94,gol95,kot00}
and Nagorny \etal\ \cite{nag94a,nag94b}.
The results of Van Meijgaard and Tjon are based on
solutions of the Faddeev equations for the three-body system,
employing a central local $NN$-interaction,
the spin-dependent Malfliet-Tjon I-III potential~\cite{mal69},
in the Unitary Pole Expansion (UPE).
Since this interaction contains only $s$-wave forces,
the ground-state wave function of $^3$He includes only $s$-waves.
Furthermore, a relativistic current operator is used.

The calculations of Golak \etal\ are also  based on the solution of
Faddeev-like equations, but employ the one-boson-exchange Bonn-B
potential~\cite{mac89} and the AV18 interaction.~\cite{av18}.
Thus state-of-the-art \hedrie\ and
3N continuum wave functions are employed. A non-relativistic
single-nucleon current operator is used, consistent with the
non-relativistic Schr\"odinger equation for the 3N states.
(The lowest-order relativistic corrections to the single-nucleon
density operator were studied and found to be on the
$5\%$ level in the range of small $p_m$ values.)
In addition $\pi$- and $\rho$-like exchange currents~\cite{car98}
 are included, which in case of the AV18 potential are consistent
with the forces. The Riska prescription~\cite{ris85} was employed,
which guarantees that the currents fulfill the continuity equation.
In this approach the transverse currents $j_x$ and $j_y$ explicitly
include MEC, while $j_z$ is related to the charge current $j_0$
through the continuity equation.
MEC were not included in the charge density operator.
This would be a relativistic effect, which was not considered.
In case of the Bonn-B potential standard~\cite{mat89}
$\pi$- and $\rho$-like exchange currents are taken.
The electromagnetic nucleon form factors used are the ones from
Hoehler \etal~\cite{hoe95}.

Nagorny \etal\ use quite a different approach,
which is based on including the electromagnetic field
into the strongly-interacting system
in a fully relativistic and gauge-invariant way \cite{nag89}.
Two covariant sets of diagrams, including
pole, ``contact'' and one-loop diagrams are used,
which provide both nuclear-current conservation and inclusion of the
dominant part of FSI and MEC effects
in a form that is  consistent with the nuclear dynamics.
The diagrams were generated by ``minimal insertion'' of the electromagnetic
field into all external/internal lines and also directly into the
3- and 4-point nuclear vertices, which produces various contact currents
in accordance with Ward-Takahashi identities \cite{nag91}.
The strong form factors in the covariant nuclear vertices
$^3$He$\rightarrow pd$ and
$^3$He$\rightarrow ppn$
are taken as the positive-energy states in the laboratory frame
through the solutions of
the Faddeev equations with the AV18 potential.
All angular momentum states in the ground state of \hedrie\ are included.
The electromagnetic form factors in the completely relativistic
$\gamma NN$-, $\gamma dd$- and $\gamma^3$He$^3$He-vertices
are taken from standard parameterizations of experimental form factors.


\section{Results}

\subsection{Direct proton  knock-out}

\begin{figure}[hbt]
        \centerline{
        \psfig{figure=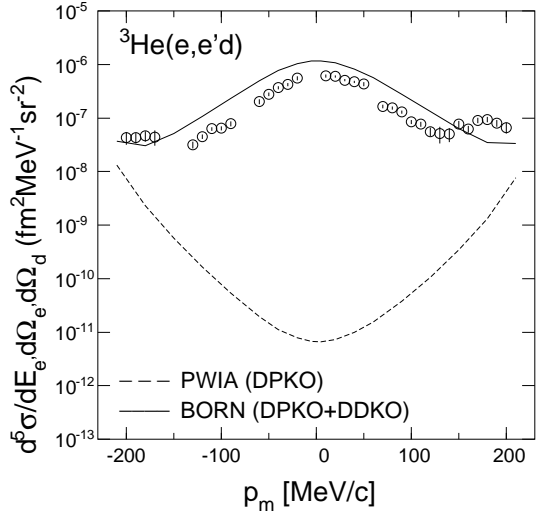,width=7.0cm}}
        \caption{
                Measured cross sections for the \hedrieeed\ reaction
                for  $q=412$~\mevc\ at $E_e=576$~MeV.
                The curves are the PWIA (dashed line) and BORN (full line)
                calculations of Tjon \etal ,
                showing the competition between DPKO and DDKO.
                }
        \label{fig:xs_412}
\end{figure}

We first focus on the measurements at $q=412$~\mevc\  at the higher beam
energy, where the largest range in $p_m$ has been probed.
The measured  \hedrieeed\ cross sections for this kinematics
are shown in Fig.~\ref{fig:xs_412}.
Above a certain value of $p_m$ the cross section flattens off,
an effect that is not seen in parallel kinematics \cite{spa98}.
The origin of this effect can easily be understood.
If one neglects FSI the virtual photon can either couple
to the proton that is knocked out,
while the detected deuteron acts as a spectator
(direct proton knock-out: DPKO),
or the virtual photon couples to either one of the nucleons that
constitute the final deuteron, which is knocked out, while
the proton is a spectator (direct deuteron knock-out: DDKO).
Inclusion of FSI modifies this picture quantitatively, but not qualitatively.

Now one can look into the kinematics.
In the case of DDKO, $p_m$ and (the constant value of) $q$
combine to the detected deuteron momentum $p_d$, which decreases
from about 400~\mevc\ at $p_m=0$ \mevc\ to about 340~\mevc\ at $p_m=200$~\mevc.
However, in case of DPKO the effective $p_m$ is equal to $p_d$.
Since for both processes the cross section decreases with the effective
value of $p_m$, the cross section is dominated at low $p_m$ by DDKO, but for
increasing $p_m$ values the contribution of DPKO increases.

Although for a quantitative comparison with the data the effects of
distortions have to be included,
the PWIA and BORN calculations of Van Meijgaard and Tjon show this nicely.
The PWIA calculation contains only the DPKO term, while the BORN calculation
includes both the DPKO and  the DDKO contribution.
Whereas for $p_m=0$ the PWIA calculation is orders of magnitude
below the data, it increases rapidly with $p_m$,
ending up only one order of magnitude below the data
at the largest values of $p_m$ measured.
The BORN calculation is largest around $p_m=0$ and then decreases,
but flattens off at large $p_m$ values due to the increasing contribution
of the DPKO part.


\subsection{Cross sections}

\begin{figure}[thb]
        \centerline{
        \psfig{figure=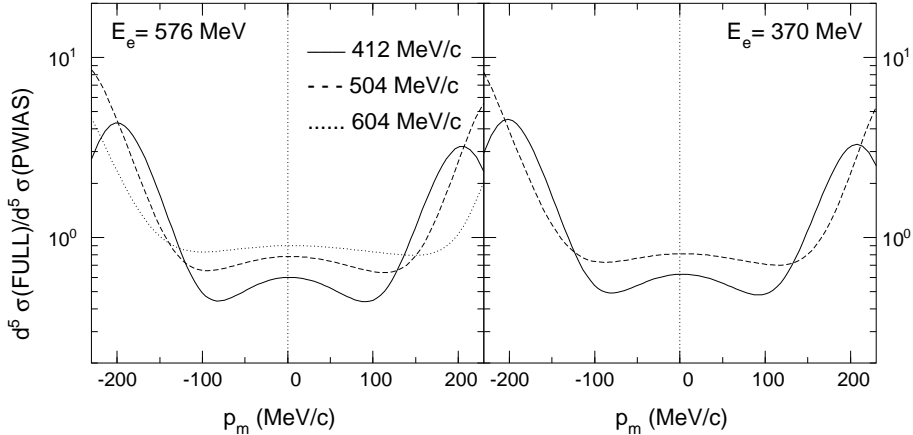,width=12.0cm}}
        \caption{Ratio of FULL and PWIAS cross sections as calculated
        by Golak \etal\ with the AV18 potential, for different values
        of the momentum transfer.
        }
        \label{fig:FSI}
\end{figure}

Before discussing all data and the comparison with the calculations, it
is of interest to look into the influence of final-state interactions (FSI).
The ratio of the FULL cross sections, i.e., including FSI, over the ones
calculated in PWIAS, which means a fully antisymmetrized plane-wave
impulse approximation \cite{gol95} (equivalent to the BORN cross section
by Van Meijgaard and Tjon), is shown in Fig.~\ref{fig:FSI}.
One observes that FSI effects around $p_m=0$ decrease the cross
section considerably at $q=412$~\mevc, but less so at larger $q$.
This has already been noticed before~\cite{gol95}. Here,
one should keep in mind that the $p-d$ centre-of-mass energy is only
14.7~MeV at the lowest $q$-value, rising to 31.1~MeV at the highest
$q$-value. In the latter case the reduction of the cross section
at low values of $p_m$ is already rather small, about 10-15\%.
Beyond a $|p_m|$-value of
about 150 \mevc\ FSI increases the cross section considerably.
(Overall the effects are slightly smaller at the lower electron energy.
We will come back to this later when discussing the separated
structure functions).

\begin{figure}[thb]
        \centerline{
        \psfig{figure=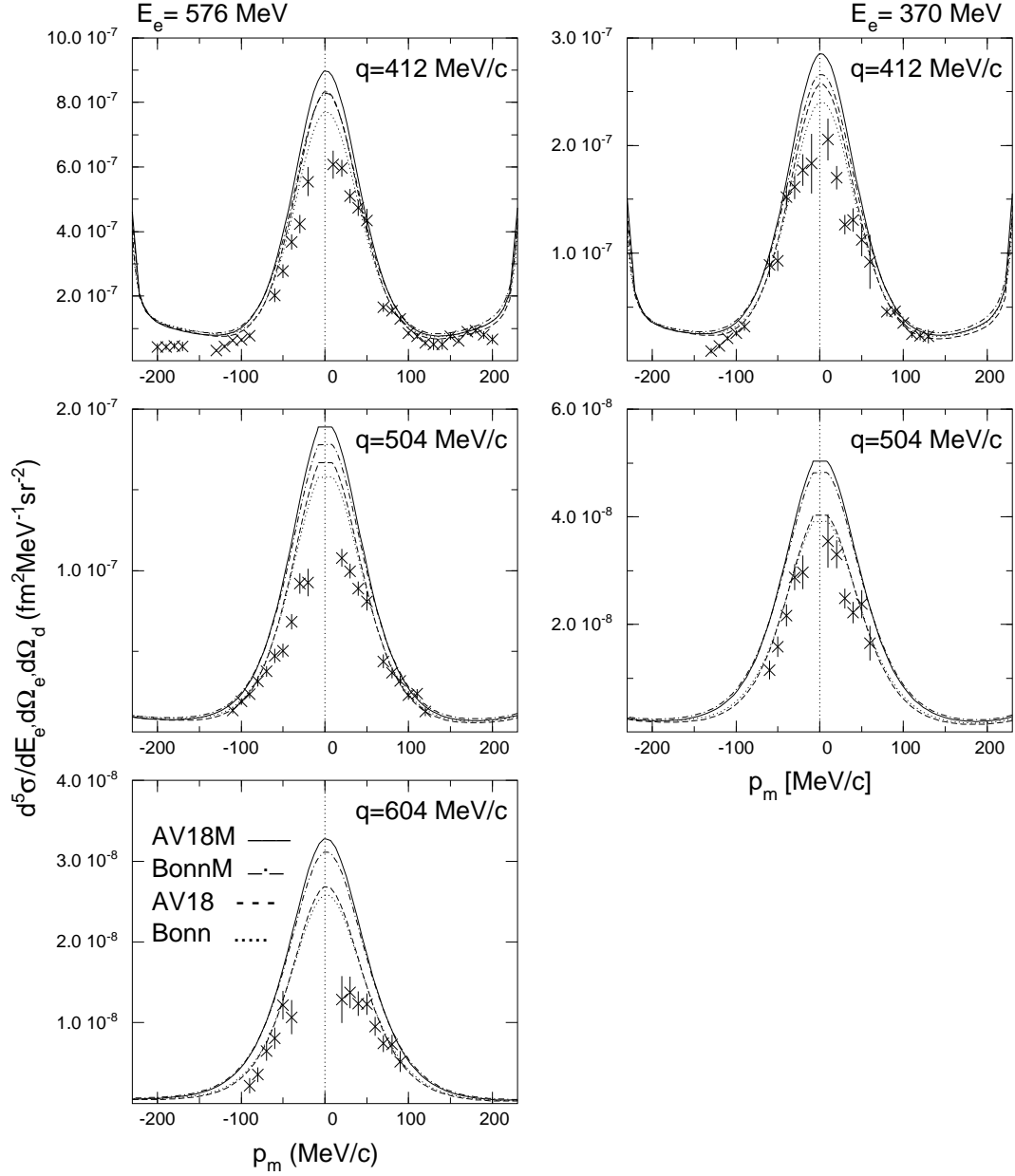,width=14.0cm}}
        \caption{Measured cross sections for different values of $q$ at
        the two beam energies, compared to the results of the calculations by
        Golak \etal\ for the Bonn-B (dotted and dash-dotted curves)
        and AV18 potential (dashed and full curves),
        without and with the inclusion of MEC, respectively.
        }
        \label{fig:xs_golak}
\end{figure}

The measured \eed\ cross sections for the various kinematics
are shown in Fig.~\ref{fig:xs_golak}.
Since the calculations of Golak \etal\ are the most detailed,
we will first compare those to the data.  One can make three observations.
The first one is that the calculations overestimate the data. This holds
already for the results with the Bonn-B  potential, and even more
for the AV18 ones. The inclusion of MEC increases the discrepancy.
Secondly, the (relative) difference between
data and calculations is slightly smaller at the lower beam energy
(we will discuss this further in the next subsection)
and increases  with $q$.
Finally, the $p_m$~dependence of the data is fairly well described, except
at $q=412$~\mevc, where there is a discrepancy at negative $p_m$,
in the region where the cross section flattens off due to the influence of DPKO.
Obviously the interference between the DPKO and DDKO processes is not
well described.

\begin{figure}[thb]
        \centerline{
        \psfig{figure=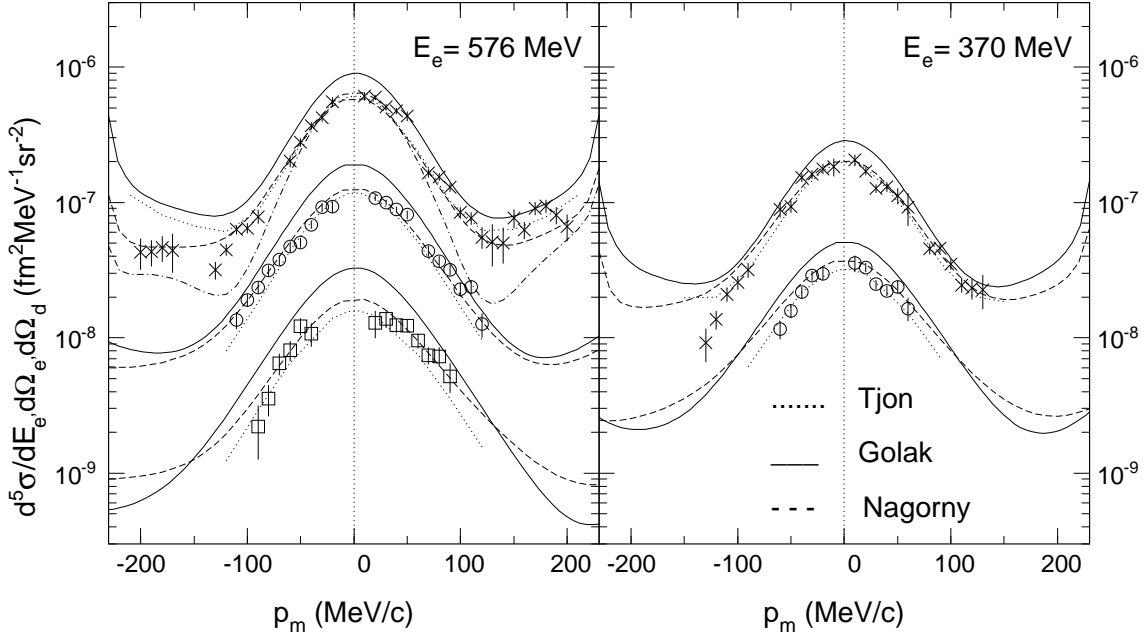,width=15.0cm}}
        \caption{
        Measured cross sections at both electron energies
        for the different values of $q$:
        412~\mevc\ (crosses), 504~\mevc\ (circles) and 604~\mevc\ (squares).
        The data is shown together with the FULL calculations
        of Van Meijgaard and Tjon (dotted line),
        Golak \etal\  (full line) and
        Nagorny (dashed line). See the text for the dash-dotted line.
        }
        \label{fig:xs_all}
\end{figure}

In Fig.~\ref{fig:xs_all} the three theoretical approaches are compared with
each other and with the data. For the calculations by Golak \etal\ the
AV18 results are chosen, since these forces and the accompanying MEC
are considered to be the most realistic ones.
In addition to what has been said already about the various calculations
some more points are worth mentioning.

It is seen, especially at small $p_m$-values, that
with increasing values of $q$ the calculations by
Van Meijgaard and Tjon predict increasingly smaller values for the
cross section compared to those calculated by Golak \etal .
Since in first approximation the $q$-dependence
of the cross section probes the wave function of the $pn$-pair,
part of this may reflect the increasing importance of the $D$-state
of the $pn$-pair (compare elastic scattering from the deuteron
at these $q$-values \cite{for96}). As mentioned, the calculations by
Van Meijgaard and Tjon do not contain $d$-waves.
Golak \etal\ have also performed calculations with the MT I-III potential,
using their non-relativistic single-nucleon current.
The results agree quite well with the ones by Van Meijgaard and Tjon.

The calculations by Nagorny give a fairly good description of the data.
At $q$~=~412~\mevc\ earlier calculations \cite{thesis}
were significantly lower than the data for  $p_m>$100~\mevc\ (see
the dash-dotted line in Fig.~\ref{fig:xs_all}).
In contrast to those the present calculations include also the $l=2, L=2$
configuration in the \hedrie\ ground state. Although this component is
small, it has a relatively large influence at larger values of $p_m$
due to its overlap with the deuteron $l=2$ (D) state.
Altogether, these calculations give a fairly good description of both
the $p_m$- and the $q$-dependence of the data.

\subsection{Structure functions}

As mentioned in section~\ref{sect:exp}, the experimental data was obtained
for values of $\phi$ close to 0 and $\pi$, and for two values of the
incoming electron energy, so the structure functions
$W_L+{v_{TT} \over v_L}W_{TT}$, $W_T$ and
$W_{LT}$ (see Eq.~\ref{eq:xs_in_vW}) could be separated.

\begin{figure}[thb]
        \centerline{
        \psfig{figure=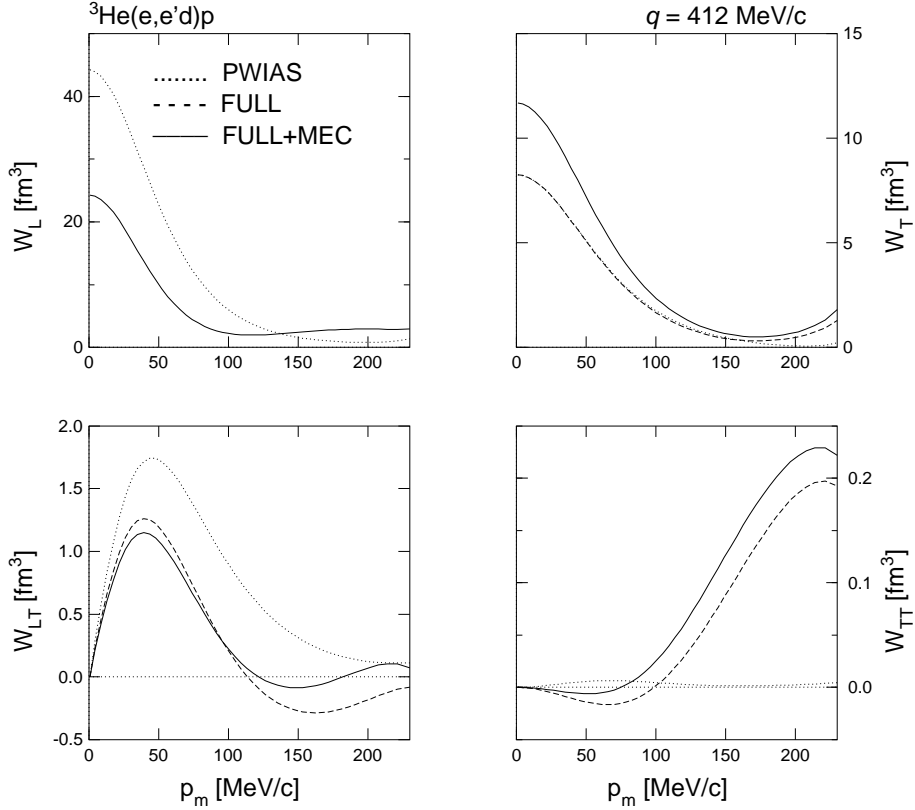,width=12.0cm}}
        \caption{Structure functions calculated by Golak \etal\ with
        the AV18 potential for $q=412$ \mevc\,.
        Notice the different vertical scales for the different W's.
        }
        \label{fig:Wxy_golak_412}
\end{figure}

Before discussing the data it is instructive to look at the theoretical
predictions. In Fig.~\ref{fig:Wxy_golak_412} the separate structure
functions at $q=412$~\mevc\ calculated by Golak  with the
AV18~interaction, with and without inclusion of FSI and MEC, are shown.
It is clear that $W_L$ and $W_T$ are the dominant structure functions,
and that $W_T$ is not small compared to $W_L$.
This is due to the contribution of the (fully transverse)
$\Delta T=1$ transition (the $\Delta T=0$ transition is dominantly
longitudinal), see Refs.~\cite{tri96,spa99}.
The $W_{LT}$ interference structure function is rather small,
while $W_{TT}$ is negligible except for values of $p_m$ above about 150~\mevc.
\newline
Another interesting observation is that the influence of FSI
is large in $W_L$ (and in $W_{LT}$ and $W_{TT}$),
but very small in $W_T$, except at large values of $p_m$.
In this context one should realize that a PWIAS calculation
for $W_L$ violates current conservation.
On the other hand MEC, as included in the present approach,
only influence the 'transverse current' dependent $W$'s,
considerably increasing the values of $W_T$ and $W_{TT}$.

$W_{LT}$

\begin{figure}[thb]
        \centerline{
        \psfig{figure=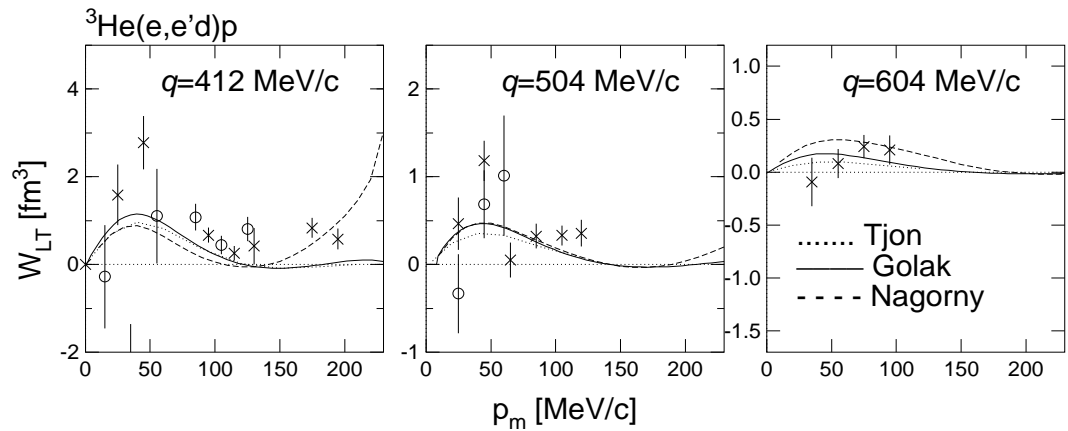,width=14.0cm}}
        \caption{
        Measured $W_{LT}$ structure functions, extracted from the
        cross sections measured at $E_e$=576~MeV (crosses) and
        $E_e$=370~MeV (circles), for the different $q$ values,
        in comparison to the calculations.
        }
        \label{fig:wlt}
\end{figure}

The $W_{LT}$ structure function can be determined by just comparing
the data in kinematics with $\theta_d$ smaller and larger than $\theta_q$
(see section~\ref{sect:exp}).
The separated $W_{LT}$ structure functions for the three values of $q$
are shown together with the calculations in Fig.~\ref{fig:wlt},
both for $E_e$=~576~MeV and $E_e$=~370~MeV.
The values obtained at the two different energies agree within the
experimental uncertainties.
As is already clear on the cross section level,
where the difference between the data for $\theta_d$ smaller
and larger than $\theta_q$ is not very large,
the measured $W_{LT}$ is small, so the error bars are relatively large.
The theoretical curves for $W_{LT}$ were obtained by separating the
calculated cross sections in the same way as the measured cross sections.
(Naturally they agreed with the ones calculated directly, using the
appropriate components of the hadronic current operator).
For all calculations $W_{LT}$ is independent of the incoming energy,
as expected, and has a maximum near $p_m$=~50~\mevc\ for all $q$ values,
while it decreases for higher $p_m$ values.
In the Faddeev calculations FSI has two effects on $W_{LT}$:
both the magnitude of the maximum decreases and the decrease
as a function of $p_m$ is stronger.
This even introduces a zero crossing at $q$=~504~\mevc\
and a double zero crossing at $q$=~412~\mevc .
The latter presumably is related to the increasing influence of DPKO,
which is also visible in this region in the cross section.
\newline
In the $p_m$ range between 0 and about 70 \mevc\ the data and the various
calculations agree with each other within the experimental accuracy.
However, at larger values of $p_m$ the calculations tend to underestimate
the data.

$W_L$ and $W_T$

\begin{figure}[htb]
        \centerline{
        \psfig{figure=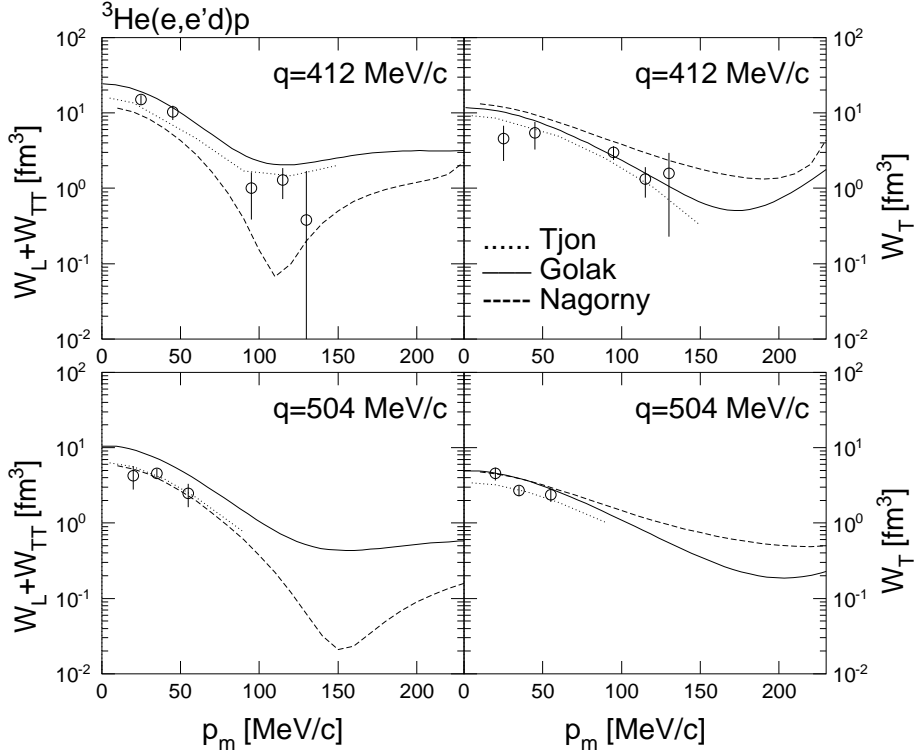,width=12.0cm}}
        \caption{
        Data for the structure functions $W_L\,(+W_{TT})$
        and $W_T$ together with the results of the calculations.
        }
        \label{fig:wl_wt}
\end{figure}

Data at  $q$=~412 \mevc\ and $q$=~504~\mevc\ were taken at both energies.
Therefore, for the first time the $W_L(+{v_{TT} \over v_L}W_{TT})$
and $W_T$ structure functions could be separated in \qw\ kinematics,
for those $p_m$ values where all four cross sections were available.
Since all calculations predict $W_{TT}$ to be very small compared to
$W_L$, except at values of $p_m$ above about 150 \mevc,
this is effectively an L-T separation.
The results are presented in Fig.~\ref{fig:wl_wt}.
It turns out that the values found for $W_L$ and $W_T$ in \qw\ kinematics
are very close to the values found in parallel kinematics~\cite{spa98}.
This, together with the fact that $W_{LT}$ and $W_{TT}$ are small,
suggests that a factorization of the cross section into an '$e-pn$'
cross section and a (distorted) spectral function (see also the introduction
and Ref.~\cite{ent94}) is a reasonable approximation.

The calculations by Golak \etal\ with the AV18 potential and including MEC
are in global agreement with $W_T$, but overestimate $W_L$, especially
at $q=504$~\mevc.  In this context we recall that at $q=604$~\mevc,
where no separation of the structure functions could be performed,
the unseparated cross section was even more overestimated.
Since the $q$-dependence in first instance probes the relative motion
of the $pn$ pair in \hedrie, this suggests that the short-distance
behaviour of this motion is not well described.
(As the used current operator is a non-relativistic one,
it was verified that elastic scattering from the free deuteron
was well described).

Whereas the calculations by Nagorny gave a fairly  good description of
the measured cross sections, they clearly underestimate the longitudinal
structure function $W_L$ at $q=412$~\mevc, and overestimate $W_T$.
The calculations show a minimum for $W_L$, which
is not present in a PWIAS calculation.
This has been observed before by Nagorny \etal\ \cite{nag94a}.
The cause is a large, destructively interfering, contribution
of the $^3$He-pole diagram (S-term) in \qw\ kinematics.
This term, which is part of the FSI, is constant with $p_m$,
since it only depends on $\omega$.
Another part of the FSI comes from the ``contact current'', but this part
is still small at $q=412$~\mevc.
Clearly, this indicates an inadequacy of the treatment of FSI.
Presumably, inclusion of only the pole part of
$pd \rightarrow pd$ rescattering is not sufficient,
and  the regular part \cite{nag91} has to be included as well.

Although the pioneering calculations by Van Meijgaard and Tjon give a
reasonable description of the data, this observation does not allow
strong conclusions, since the used interaction is rather simple,
no MEC are included, and only $s$-waves are taken into account.

\section{Summary and conclusions}

Cross sections for the  \hedrieeedp\ reaction have been measured
in \qw -constant kinematics for a range in missing momentum
at two beam energies and for three values of the three-momentum transfer.
Thus the separate L(+TT), T and LT structure functions
could be determined at $q$~=~412 and 504~\mevc, which provide for
a sensitive test of theoretical calculations for this reaction.
The LT structure function is found to be rather small, but the
transverse structure function is of comparable magnitude as the
longitudinal one. This points to the importance of the $\Delta T=1$
transition, in which a $T=1$ \pn\ pair in \hedrie\ is transformed
into a $T=0$ deuteron in the final state.

The data has been compared to the results of two types of
three-body Faddeev calculations, one with a simple interaction
and containing only $s$-waves, the other one employing the Bonn-B
and the AV18 potentials, and including meson-exchange currents (MEC).
Also calculations based on a covariant diagrammatic approach,
including tree and one-loop diagrams, and using Ward-Takahashi
identities to ensure gauge invariance were used.

All calculations give a fair description of the $p_m$- and the
$q$-dependences of the cross sections, which reflect the centre-of-mass and
relative motion of the \pn\ pair inside \hedrie, respectively,
and of their L/T character.
This means that the essential ingredients of the calculations: the
structure of \hedrie\ in its ground state, and final-state interactions
(the continuum structure of the 3N system), are reasonably well
understood.
Final-state interactions have a large influence, especially
at the lower value of $q$ (and accompanying low centre-of-mass
energy of the final $p$-$d$ system).
At low values of $p_m$ they reduce the longitudinal part of the
cross section, whereas at large $p_m$-values they increase it.
In this region also direct proton knock-out starts to noticeably
influence the cross sections, leading to a rise of the cross section.
This was experimentally observed at $p_m$-values above 150 \mevc\ at
the lowest value of $q$.
In contrast the value of the transverse structure function hardly changes
when including FSI.
Meson-exchange currents increase considerably the
transverse structure function.
All calculations predict relatively small values for the LT
interference structure function, which is consistent with the data.
Also $W_{TT}$ is calculated to be small, but this structure function
cannot be separated with the present ``in-plane'' data.

However, upon closer inspection there are also significant differences
between the data and the theoretical calculations.
The diagrammatic approach gives a fair description of the cross section,
but under(over)estimates the longitudinal (transverse) structure function.
It also predicts a minimum in the longitudinal structure function,
presumably due to the neglect of the regular part of $pd \rightarrow pd$
rescattering.
The most striking result is that the supposedly best Faddeev calculations
available at present, which employ the AV18 nucleon-nucleon interaction
and include MEC, overestimate the measured cross sections,
the more the larger the value of $q$.
The major discrepancy is in the longitudinal cross section,
the transverse one being reasonably well described.
Possible explanations for this discrepancy could be:
the use of the Sachs form factors $G_E^p$ and $G_E^n$ instead of 
the Pauli form factors $F_1^p$ and $F_1^n$ reduces the
longitudinal response function $R_L$ by about 15-25\%,
depending on the kinematics. Furthermore, inclusion of a
three-nucleon force will influence the binding energy and hence the
$d-p$ overlap. A PWIAS calculation including the Urbana IX 3N
force \cite{Pud97} for the \hedrie\ bound state gives a reduction
of the cross section of about 15\%. Finally meson-exchange currents should also
be included in the charge-density operator, since it is known
that they influence e.g. the elastic charge form factor. However,
such calculations are not yet available.

\begin{ack}
This work is part of the research programme of the ``Stichting voor
Fundamenteel Onderzoek der Materie (FOM)'', which is financially
supported by the ``Ne\-der\-land\-se Organisatie voor Wetenschappelijk
Onderzoek (NWO)''.
The Faddeev calculations by Golak \etal\ were supported by the Deutsche
Forschungsgemeinschaft and the Polish Committee for Scientific Research.
The calculations were performed on the T90 of the NIC in J\"{u}lich.
\end{ack}

\end{document}